# Max-Leaves Spanning Tree is APX-hard for Cubic Graphs

Paul Bonsma

Humboldt Universität zu Berlin, Computer Science Department,
Unter den Linden 6, 10099 Berlin.

May 31, 2018

**Abstract**

We consider the problem of finding a spanning tree with maximum number of leaves (MaxLeaf). A 2-approximation algorithm is known for this problem, and a 3/2-approximation algorithm when restricted to graphs where every vertex has degree 3 (*cubic* graphs). MaxLeaf is known to be APX-hard in general, and NP-hard for cubic graphs. We show that the problem is also APX-hard for cubic graphs. The APX-hardness of the related problem Minimum Connected Dominating Set for cubic graphs follows.

## 1 Introduction

We study the problem Maximum Leaf Spanning Tree or MaxLeaf, for which the objective is to find in a given connected graph a spanning tree with maximum number of leaves. An *α-approximation algorithm* for a maximization (minimization) problem is a polynomial time algorithm that returns a solution with objective value at least (at most) $\alpha \cdot \text{OPT}$, where OPT is the objective value of an optimal solution for the given instance[1]. MaxLeaf is known to be APX-hard [12], which implies that there exists an $\epsilon > 0$ such that no polynomial time $(1-\epsilon)$-approximation algorithm is possible for this problem, unless P=NP [2]. However, constant factor approximation algorithms are known: Lu and Ravi [20] gave a 1/3-approximation, and this was later improved by Solis-Oba who gave a linear time 1/2-approximation [23]. So the problem is in APX – the class of optimization problems with constant factor approximation algorithms – and therefore APX-complete.

MaxLeaf is closely related to *Minimum Connected Dominating Set* (MinCDS). This problem asks, given a graph $G$, for a set $S \subset V(G)$ of minimum size such that $G[S]$ is connected and every vertex $v \notin S$ is adjacent to a vertex in $S$ (a *connected dominating set*). The relation between these problems is as follows: since the non-leaves of a spanning tree of $G$ form a connected dominating set (unless $G = K_2$), $G$ has a spanning tree with at least $k$ leaves if and only if $G$ has a connected dominating set of size at most $|V(G)| - k$. These problems differ from an approximability viewpoint: Guha and Khuller [14] showed that MinCDS admits no constant factor approximation algorithm under established complexity-theoretic assumptions. Ruan et al [22] give a $2 + \ln \Delta(G)$-approximation algorithm, where $\Delta(G)$ is the maximum degree of $G$.

In *cubic* graphs, every vertex has degree 3. The restriction of MaxLeaf to cubic graphs has received much attention. One reason is that these are easier to analyze algorithmically, yet from an approximation viewpoint, this is where the main hardness lies. For instance, for 5-regular graphs a 2/3-approximation follows easily from known bounds [13], see below. For cubic graphs, more work is required to obtain this ratio: Loryś and Zwoźniak [18] gave a 4/7-approximation for MaxLeaf on cubic graphs. This ratio was later improved to 3/5 by Correa et al [6], and finally by Bonsma and Zickfeld [4] to 2/3. A natural question is how far this can be improved. However, even the question whether the problem is APX-hard for cubic graphs remained open. This question was asked in [6] and [4]. The only known hardness result for cubic graphs is that the problem is NP-hard, as was shown by Lemke in an unpublished technical report [17].

---

[1]In the literature on MaxLeaf, approximation algorithms are usually stated with $\alpha > 1$ approximation ratios. For our proofs it is more convenient to define these as $1/\alpha$-approximation algorithms.

In this paper we settle the question by showing that also for cubic graphs, the problem is APX-hard. This is strictly stronger than the known hardness results [17, 12]. From this the APX-hardness of MinCDS for cubic graphs will also follow. The proof is interesting by itself, since it shows how APX-hardness results can be proved using extremal arguments. Informally speaking, the problem with proving APX-hardness for cubic graphs is that it seems impossible to find 'well-behaved' gadgets, that allow for an easy analysis of the graph constructed in the reduction. Instead we have a simple construction, but need an elaborate global analysis of the constructed graph, involving various (fractional) bounds and rounding arguments. As a contrast, we give a new very simple and more traditional APX-hardness proof for MaxLeaf in general graphs in at the end of this introduction.

APX-hardness results for basic problems in restricted graph classes, in particular cubic graphs, are useful since they allow for simple hardness proofs of many other problems. The four hardness results by Alimonti and Kann [1] have often been used for this purpose: they show that the problems Minimum Vertex Cover, Maximum Independent Set, Minimum Dominating Set and Maximum Cut are APX-hard when restricted to cubic graphs. Their APX-hardness results for Maximum Independent Set and Minimum Vertex Cover will be used for the two reductions in this paper.

We now review some algorithmic results on MaxLeaf. Recently, the generalization of MaxLeaf to directed graphs or *digraphs* has received a lot of attention. Very recently Daligault and Thomasse [7] gave a constant factor approximation algorithm for this problem (more precisely, a 1/92-approximation algorithm), improving on the $\Omega(1/\sqrt{\text{OPT}})$-approximation of Drescher and Vetta [9]. The paper of Daligault and Thomasse [7] also deals with the parameterized variant of the decision version of Directed MaxLeaf. See [10, 16] for other parameterized results on (un)directed MaxLeaf. Undirected MaxLeaf has also been studied in the area of fast exact algorithms. Fomin et al [11] gave an algorithm for finding a minimum connected dominating set, and therefore a maximum leaf spanning tree, that runs in time $O(1.9407^n)$ where $n$ is the number of vertices.

Combinatorial bounds form an important ingredient for many of the above results. For instance, it is known that connected graphs with minimum degree $\delta \geq 3$ on $n$ vertices admit a spanning tree with at least $n/4 + 2$ leaves [15]. A stronger version of this bound appears in [5]. For cubic graphs, see [4] for an improved bound. When $\delta \geq 4$, $2n/5 + 8/5$ leaves are possible [15, 13], and for $\delta \geq 5$, $n/2 + 2$ leaves are possible [13]. In [3] and [7] bounds for the directed case can be found.

One may wonder why it is much harder to prove APX-hardness for cubic graphs than it is to prove NP-hardness for cubic graphs [17] or APX-hardness for general graphs [12]. Indeed, for general graphs a very simple APX-hardness proof can be given, using a reduction from the APX-hard problem Cubic Minimum Vertex Cover: let $G$ be a cubic graph on $n$ vertices and $m = \frac{3}{2}n$ edges for which we search a *minimum vertex cover*, i.e. a minimum set $S \subseteq V(G)$ such that every edge of $G$ is incident with some vertex of $S$. Let $k$ be the size of a minimum vertex cover. Construct a MaxLeaf instance $G'$ as follows: introduce a new vertex $x$, and add edges from $x$ to every other vertex. Next, subdivide every edge not incident with $x$ with a single vertex. It can be checked that any spanning tree in $G'$ can be transformed into a spanning tree with at least as many leaves, where all the degree 2 vertices are leaves. From this it follows that $G$ has a vertex cover with at most $y$ vertices if and only if $G'$ has a spanning tree with at least $n - y + m$ leaves. Since $G$ is cubic, $k \geq m/3$. A $(1-\epsilon)$-approximation algorithm for MaxLeaf now yields a solution with at least $(1-\epsilon)(n-k+m) = n-k+m-\epsilon(5/3m-k) \geq n-k+m-\epsilon(5k-k) = n-(1+4\epsilon)k+m$ leaves, and therefore a vertex cover of size at most $(1+4\epsilon)k$. This concludes the APX-hardness proof.

It seems however impossible to give a similar simple proof for cubic graphs. Considering the NP-hardness proof for cubic graphs, Lemke [17] gave a reduction from *Exact Cover by 3-Sets*. Here a 3-uniform hypergraph $G$ on $n$ vertices is given (i.e. all edges contain three vertices). The question is whether there is a subset of the edges $Q$ such that every vertex is contained in exactly one edge of $Q$. For every instance $G$, in [17] a graph is constructed that has a spanning tree without vertices of degree 2 if and only if $G$ is a 'yes'-instance. It is easily seen that such a tree is optimal. However, an approximation preserving reduction from an APX-hard problem needs also to take into account cases where the tree is not optimal, that is, it contains some degree 2 vertices. In this case the behavior of the subgraphs in Lemke's construction, or even any cubic construction, becomes much harder to analyze.

In Section 2 we give definitions and notations, and in Section 3 the construction of our APX-hardness proof, which uses an approximation preserving reduction from Cubic Maximum Independent Set. Sections 4 and 5 show how leafy spanning trees yield large independent sets and vice versa, and in Section 6 these bounds are combined to conclude the proof.

## 2 Preliminaries

For basic graph theoretic definitions, we follow [8]. By $d_G(v)$ we denote the degree of $v$ in graph $G$. The subscript is omitted when the graph in question is clear. By $\delta(G)$ and $\Delta(G)$ we denote the minimum and maximum degree of $G$, respectively. By $G - S$ we denote the graph obtained from $G$ by deleting the vertex or edge set $S$.

A directed graph or *digraph* $D$ consists of a vertex set $V(D)$ and arc set $A(D)$, which is a set of ordered 2-tuples of vertices. For an arc $(u,v) \in A(D)$, $u$ is called the *tail* and $v$ the *head* of $(u,v)$. The *in-degree* $d^-(v)$ (*out-degree* $d^+(v)$) of a vertex $v$ is the number of arcs of which $v$ is the head (tail). A directed graph (*digraph*) $D$ is an *orientation* of an undirected graph $G$ if $V(D) = V(G)$ and there exists a bijection $f : A(D) \to E(G)$ with $f((u,v)) = \{u,v\}$ for all $(u,v) \in A(D)$. An *out-tree orientation* of a tree $T$ is an orientation $T'$ of the (given undirected) tree $T$ such that $T'$ is an *out-tree*, that is, there is exactly one vertex with in-degree 0, which is called the *root*. Note that every other vertex then has in-degree 1.

A vertex sequence $v_0, \ldots, v_k$ is called a path or cycle in a digraph $D$ if it is a path or cycle in the underlying undirected graph (i.e. $(v_i, v_{i+1}) \in A(D)$ or $(v_{i+1}, v_i) \in A(D)$ holds for all $i$). Directed paths and cycles, where $(v_i, v_{i+1}) \in A(D)$ holds for all $i$ are called *dipaths* and *dicycles*. A path from $u$ to $v$ is also called a $(u,v)$-*path*. In an undirected graph $G$, $v$ is said to be *reachable* from $u$ if a $(u,v)$-path exists in $G$. In a digraph $D$, $v$ is reachable from $u$ if a $(u,v)$-dipath exists.

An induced subgraph $H$ of an undirected graph $G$ is called a $k$-*terminal subgraph* if $H$

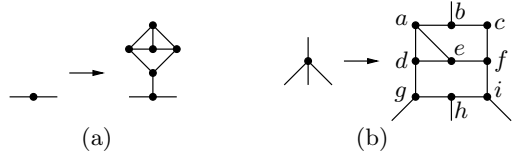

(a)　(b)

Figure 1: Gadgets used in the construction.

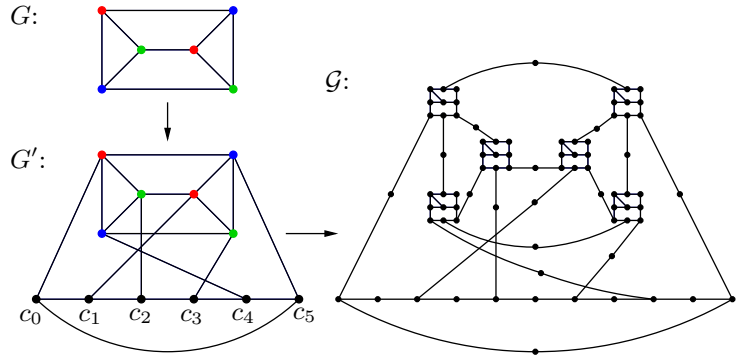

Figure 2: Constructing a Weighted MaxLeaf instance from a Cubic MIS Instance.

contains exactly $k$ vertices that have neighbors outside of $H$, these are called its *terminals*.

## 3  The Construction of a Weighted MaxLeaf Instance

We now prove that Cubic MaxLeaf is APX-hard (and thus APX-complete), using a reduction from *Cubic Maximum Independent Set (Cubic MIS)*. This problem has as input a cubic graph $G$, and asks for a maximum size set $S \subseteq V(G)$ such that no two vertices in $S$ are adjacent. To improve the presentation, we will prove that the following problem variant is APX-hard, from which APX-hardness of cubic MaxLeaf easily follows. The problem *Weighted MaxLeaf* has as input a graph $G$ with $\Delta(G) \leq 3$ and $\delta(G) \geq 2$, and the objective is to find a spanning tree $T$ that maximizes the number of vertices $v$ with $d_T(v) = 1$ and $d_G(v) = 3$. We will also call vertices of $G$ with degree 3 *weighted vertices* and the other vertices *unweighted*. So the objective is to maximize the number of *weighted leaves*. By $\ell(T)$ we will denote the number of weighted leaves of $T$.   $\ell(T)$

From instances $G$ of Weighted MaxLeaf, it is easy to construct equivalent Cubic MaxLeaf instances: replace every vertex of degree 2 by the 1-terminal subgraph as shown in Figure 1(a) (the two half edges indicate the terminal). The next lemma is easily observed.

**Lemma 1** *Let $G'$ be the cubic graph obtained from a graph $G$ with $\delta(G) = 2$, $\Delta(G) = 3$ by replacing all $x$ vertices of degree 2 as shown in Figure 1(a). Then $G'$ has a spanning tree with at least $l + 3x$ leaves if and only if $G$ has a spanning tree with at least $l$ weighted leaves.*

The construction of Weighted MaxLeaf instances uses the following gadgets. A *vertex gadget* of $G$ is an induced 4-terminal subgraph of $G$ as shown in Figure 1(b), where the four   vertex terminals are indicated by half edges. Note that one vertex has degree 2, and therefore does   gadget not count towards the number weighted leaves.

**Construction** Let $G$ be a Cubic MIS instance on $n$ vertices. We use this to construct in polynomial time a weighted MaxLeaf instance as follows. First, we assume w.l.o.g. that $G \neq K_4$, and thus we can construct a proper 3-coloring of $G$, using colors red, green and blue. (By Brooks' Theorem such a coloring exists, and in addition it can be found in polynomial time, see also [19].) Let $r$ and $g$ be the number of red and green vertices respectively, and w.l.o.g. assume $r \geq 1$ and $g \geq 1$. Number the vertices of $v_0, \ldots, v_{n-1}$ such that $v_0, \ldots, v_{r-1}$ are red, $v_r, \ldots, v_{r+g-1}$ are green, and $v_{r+g}, \ldots, v_{n-1}$ are blue. We construct a graph $\mathcal{G}$ as follows. The construction is illustrated in Figure 2.

1. Start with $G$. Add a cycle consisting of $n$ *connection vertices* $c_0, \ldots, c_{n-1}$ and edges $c_i c_{(i+1) \bmod n}$ for $i \in \{0, \ldots, n-1\}$.

2. Add edges $v_i c_i$ for all $i \in \{0, \ldots, n-1\}$.

3. Subdivide every edge with one new vertex (of degree 2).

4. Replace every vertex $v_i$ of degree four with a vertex gadget $H_i$, such that every terminal of $H_i$ becomes adjacent to a different neighbor of $v_i$. (Choose arbitrarily which terminals become adjacent to which neighbors.)

Let $\mathcal{G}$ be the resulting graph, and let $G'$ be the graph obtained after Step 2 in this construction. Recall that by our definition of Weighted MaxLeaf, the vertices introduced in Step 3 do not count towards the number of weighted leaves. For the proofs below it will be useful to denote how end vertices of edges of $G'$ correspond to vertices of $\mathcal{G}$. In Step 3, edges $uv$ of $G'$ are subdivided with a new vertex $w$ to yield two edges $uw$ and $vw$. In Step 4, the edge $uw$ may be replaced by an edge $tw$, where $t$ is a terminal of a vertex gadget. If this is the case, $t_{uv}(u)$ will denote this terminal $t$, otherwise $t_{uv}(u)$ will denote $u$.

We will proceed to show that for every $x \in \mathbb{R}$, if $\mathcal{G}$ has a spanning tree with at least $3.75n + 1.5x$ weighted leaves, then $G$ has an independent set of size at least $x - \frac{1}{3}$ (Section 4), which can be constructed in polynomial time. In addition, if $G$ has an independent set of size $x$, $\mathcal{G}$ has a spanning tree with at least $\lfloor 3.75n + 1.5x \rfloor$ weighted leaves (Section 5). In Section 6 it is then shown that this yields a $(1 - 141\epsilon)$-approximation algorithm for Cubic MIS, when a $(1 - \epsilon)$-approximation algorithm for Cubic MaxLeaf is given. This proves APX-hardness for Cubic MaxLeaf.

## 4 Constructing an Independent Set from a Spanning Tree

We first take a closer look at the behavior of vertex gadgets, by bounding the number of weighted leaves a spanning tree may contain within one given vertex gadget.

**Proposition 2** *Let $\mathcal{G}$ be a weighted MaxLeaf instance, $T$ be a spanning tree of $\mathcal{G}$ and $H$ be a vertex gadget of $\mathcal{G}$. Let $T'$ be an out-tree orientation of $T$ with root $r^* \in V(\mathcal{G}) \backslash V(H)$. Then the following bounds hold:*

(i) *$H$ contains at most six weighted leaves of $T$.*

(ii) *If $T'$ contains at least one arc leaving $V(H)$, then $H$ contains at most four weighted leaves of $T$.*

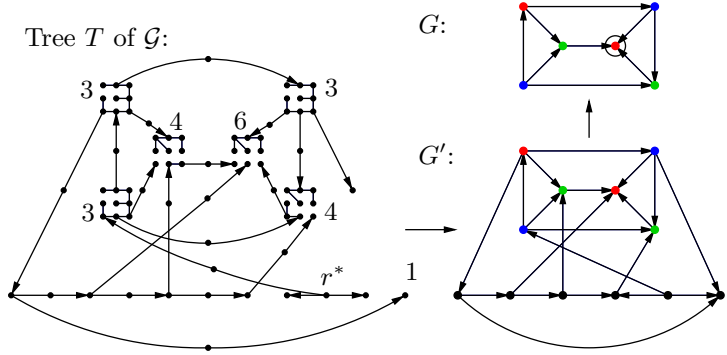

Figure 3: A spanning tree with $24 = \lceil 3.75 \cdot 6 + 1.5 \rceil$ weighted leaves yields a size 1 independent set.

(iii) *If $T'$ contains at least two arcs leaving $V(H)$, then $H$ contains at most three weighted leaves of $T$.*

*Proof:* In the proof we will refer to the vertex labels of $H$ as shown in Figure 1(b).

(i) $\{a, d, f\}$ and $\{b, g, i\}$ are vertex cuts of $\mathcal{G}$, so both contain at least one non-leaf of a *spanning* tree. They are disjoint, so $H$ contains at least two weighted non-leaves of $T$.

(ii) Since every arc of $T'$ that leaves $V(H)$ is part of a dipath in $T'$ that starts at the root, $T$ contains a path $P$ in $H$ from one terminal of $H$ to another, where all vertices of $P$ are non-leaves. Suppose $b$ is one of the ends of $P$. Then either $P$ contains at least four weighted vertices, or $P$ contains the vertices $b$, $c$, $f$ and $i$. In the second case the vertex cut $\{a, e, g\}$ shows there is at least one more non-leaf, so in both cases we have found four weighted non-leaves. Now suppose $g$ is one of the ends of $P$. If $h$ is the other end this ensures that $g$ and $h$ are non-leaves, and the two disjoint vertex cuts $\{a, d, f\}$ and $\{b, e, i\}$ show there are at least two more weighted non-leaves. If $i$ is the other end, $P$ either has length at least four (in which case we are done), or it contains $g$, $h$ and $i$. Then the vertex cut $\{a, d, f\}$ shows there must be at least one more weighted non-leaf. Finally, if $P$ goes from $h$ to $i$, the two vertex cuts $\{b, f\}$ and $\{a, e, g\}$ show that there are at least four weighted non-leaves.

(iii) Because there are at least two arcs leaving $V(H)$, in this case $T - L(T)$ contains a subgraph of $H$ of one of the following two forms: it contains a tree $T_H$ that contains at least three terminals of $H$, or it contains two paths between disjoint terminal pairs of $H$. (Note that all vertices of these subgraphs are non-leaves.) In the latter case five weighted non-leaves are easily found by considering shortest path lengths. Similarly, five non-leaves are also easily found when $\{b, g, h\} \subseteq V(T_H)$ or $\{b, g, i\} \subseteq V(T_H)$. If $\{b, h, i\} \subseteq V(T_H)$, four weighted leaves are only possible when $a$, $d$, $e$ and $g$ are leaves, but this is not possible since $\{a, e, g\}$ is a vertex cut. Finally, when $\{g, h, i\} \subseteq V(T_H)$, the three vertex cuts $\{b, f\}$, $\{b, d, e\}$ and $\{a, d, f\}$ show there are at least two additional weighted non-leaves. □

In the remainder of this section, we will prove the next lemma, which shows that an independent set $I$ of $G$ of sufficient size can be constructed when a spanning tree $T$ of $\mathcal{G}$ is given. The construction is illustrated in Figure 3. The constructed independent set consists of the single encircled vertex. Numbers indicate numbers of weighted leaves. The choice of the orientations is explained below.

The intuitive idea behind the next proof is as follows. Not too many vertex gadgets in $\mathcal{G}$ can contain six weighted leaves of a spanning tree $T$, since edges in vertex gadgets are needed to connect $T$. In particular, such vertex gadgets cannot be adjacent and thus form our independent set. With a similar more delicate argument we will also show that not all vertex gadgets can contain four leaves of $T$. How much every vertex gadget contributes to 'connecting $T$' is encoded by the out-degrees of vertices of $G'$ in the proof below. The proof of the lemma consists of a number of claims.

**Lemma 3** *Let $\mathcal{G}$ be constructed from a cubic graph $G$ on $n$ vertices as shown in Section 3. If $\mathcal{G}$ has a spanning tree $T$ with $\ell(T) \geq 3.75n + 1.5x$, then an independent set $I$ of $G$ with $|I| \geq x - \frac{1}{3}$ can be constructed in polynomial time.*

Let $T$ be a spanning tree of $\mathcal{G}$ with $\ell(T) \geq 3.75 + 1.5x$. To construct an independent set $I$ of $G$ with the desired size, we will first use $T$ to orient $G'$ and $G$. Observe that there is some connection vertex of $\mathcal{G}$ that is not a leaf of $T$. Choose $r^*$ to be such a vertex. Orient $T$ as out-tree with root $r^*$. An orientation of $G'$ can be obtained from the out-tree $T$ as follows: consider an edge $uv \in E(G')$, which was subdivided with a new vertex $w$ for constructing $\mathcal{G}$. So $uv$ corresponds to edges $t_1 w$ and $t_2 w$ of $\mathcal{G}$, with $t_1 = t_{uv}(u)$ and $t_2 = t_{uv}(v)$. $uv$ is now oriented as follows: if $(t_1, w) \in A(T)$, then choose the orientation $(u, v)$. If $(t_2, w) \in A(T)$, then choose the orientation $(v, u)$. Observe that this uniquely determines the direction of $uv$ in every case. Doing this for all edges of $G'$ yields the orientation of $G'$. Since $G$ is a subgraph of $G'$, this also yields the orientation of $G$ that we will use.

The set $I$ now consists of all vertices of $G$ that have out-degree 0. Clearly this is an independent set, and $I$ can be constructed in polynomial time. Let $n_i$ denote the number of vertices of $G$ with out-degree $i$, so $|I| = n_0$. Let $n'_i$ be the number of vertices of $G$ that have out-degree $i$ in $G'$. Observe that since $r^*$ is not part of a vertex gadget, $n'_4 = 0$. Note that $v_i$ has out-degree $d$ in $G'$ if and only if $T$ contains $d$ arcs leaving $H_i$. So Proposition 2 shows that if $v_i$ has out-degree 3 in $G'$, then $T$ has at most three weighted leaves in the vertex gadget $H_i$, etc. This yields:

$T$

$r^*$

$I$

$n_i$

$n'_i$

**Claim 1** *The number of non-connection vertices of $\mathcal{G}$ that are weighted leaves of $T$ is bounded by $6n'_0 + 4n'_1 + 3n'_2 + 3n'_3$.*

Since $T$ is an out-tree, every vertex of $T$ is reachable from the root $r^*$. Therefore every vertex of $G'$ is reachable from $r^*$ in the chosen orientation (possibly by multiple dipaths). Observe that every connection vertex that is a leaf in $T$ has out-degree 0 in $G'$. Let $z$ be the number of connection vertices of $G'$ that have an in-neighbor that is not a connection vertex.

$z$

**Claim 2** *At most $\lceil z/2 \rceil$ connection vertices of $\mathcal{G}$ are leaves in $T$.*

*Proof:* Let $c_{\sigma_1}, \ldots, c_{\sigma_k}$ be the connection vertices of $\mathcal{G}$ that are leaves in $T$, with $\sigma_i < \sigma_{i+1}$ for all $i$. All of these vertices have in-degree 3 in $G'$, which accounts for $k$ connection vertices that have an in-neighbor that is not a connection vertex.

Consider $c_{\sigma_i}$ and $c_{\sigma_{i+1}}$, for some $i$. Since these vertices have in-degree 3, they are not adjacent in $G'$. Therefore there is at least one connection vertex $c_l$ that lies between them on $C$ (that is, $\sigma_i < l < \sigma_{i+1}$). $G'$ contains a dipath $P$ from $r^*$ to $c_l$, which clearly cannot contain $c_{\sigma_i}$ or $c_{\sigma_{i+1}}$ as internal vertices. So unless $r^*$ also lies between $c_i$ and $c_{i+1}$, $P$ must contain a connection vertex between $c_{\sigma_i}$ and $c_{\sigma_{i+1}}$ that has an in-neighbor that is not a connection vertex.

Since the above argument can be applied for $k$ different pairs of connection vertices and $r^*$ lies only between one such pair, this accounts for $k-1$ additional such vertices. It follows that $z \geq 2k-1$. □

A second way to interpret the parameter $z$ is the following: there are exactly $z$ vertices with different out-degrees in $G$ and $G'$. In this case the out-degree in $G'$ is one higher. This observation yields the following inequality.

**Claim 3** $z + 3n'_0 + 2n'_1 + n'_2 = 3n_0 + 2n_1 + n_2$.

*Proof:* Let $k_i$ denote the number of vertices with out-degree $i$ in $G$ and out-degree $i+1$ in $G'$. From $n'_4 = 0$, $k_3 = 0$ follows. Vertices for which the out-degree increases this way correspond to in-neighbors of connection vertices in $G'$, so $z = k_0 + k_1 + k_2$. In addition we have that $n'_i = n_i - k_i + k_{i-1}$. Substituting these expressions yields the stated equality. □

With the above observations, we can bound the number of weighted leaves of $T$. Let $m = 1.5n$ be the number of arcs of $G$. By counting in-degrees we have $m = 3n_0 + 2n_1 + n_2$.

$$\ell(T) \leq 6n'_0 + 4n'_1 + 3n'_2 + 3n'_3 + \lceil z/2 \rceil \leq$$

$$\lceil 3n + 1.5|I| + 1.5n'_0 + n'_1 + z/2 \rceil \leq$$

$$\lceil 3n + 1.5|I| + 1.5n_0 + n_1 + 0.5n_2 \rceil \leq$$

$$\lceil 3n + 1.5|I| + 0.5m \rceil = \lceil 3.75n + 1.5|I| \rceil.$$

Here we used Claim 1; Claim 2; $n = n'_0 + n'_1 + n'_2 + n'_3$; $|I| = n_0 \geq n'_0$; $z/2 + 1.5n'_0 + n'_1 + 0.5n'_2 = 1.5n_0 + n_1 + 0.5n_2$ (Claim 3); $m = 3n_0 + 2n_1 + n_2$ and $m = 1.5n$, respectively.

So if $\ell(T) \geq 3.75n + 1.5x$, then $\lceil 3.75n + 1.5|I| \rceil \geq \ell(T) \geq 3.75n + 1.5x$. Since $G$ is a cubic graph, $n$ is even. It follows that $3.75n + 1.5|I|$ is half integral, so $3.75n + 1.5|I| + 0.5 \geq \lceil 3.75n + 1.5|I| \rceil \geq 3.75n + 1.5x$, and thus $|I| \geq x - \frac{1}{3}$. This concludes the proof of Lemma 3.

## 5 Constructing a Spanning Tree from an Independent Set

In this section we will prove the following lemma, which shows that a spanning tree $T$ with enough weighted leaves can be constructed when an independent set $I$ of $G$ is given. The proof consists of a number of claims.

The intuitive idea behind the proof is as follows. When given an independent set $I$ of $G$, we can construct a spanning tree $T$ of $\mathcal{G}$ that does not use any vertex gadget $H_i$ with $v_i \in I$ for 'connecting $T$'. For arguing that we can still make $T$ connected, we need to use the 3-coloring of $G$. We fix a connection vertex as root, and show that the red vertices can be reached from this root. This is needed to show that green vertices can be reached, which is in turn needed to show that blue vertices can be reached.

**Lemma 4** *Let $\mathcal{G}$ be constructed from a cubic graph $G$ on $n$ vertices as shown in Section 3. If $G$ has an independent set $I$ with $|I| \geq x$, then $\mathcal{G}$ has a spanning tree $T$ with $\ell(T) \geq \lfloor 3.75n + 1.5x \rfloor$.*

Throughout the proof we will refer to the vertex coloring of $G$ that was used for the construction of $\mathcal{G}$. Let $I$ be a *maximal* independent set of $G$ with $|I| \geq x$. We use this to construct $I$ a spanning tree with at least $\lfloor 3.75n + 1.5x \rfloor$ leaves as follows. The construction is illustrated in Figure 4, where $I$ is represented by encircled vertices in $G$. First, for all $v \in I$, orient all incident edges $xv$ of $G$ as $(x,v)$, so every $v \in I$ has out-degree 0. This is possible since $I$ is an independent set. For all edges that are not incident with a vertex from $I$, choose the direction from red to green, from green to blue or from red to blue, whichever applies. This yields the orientation of $G$. We extend this to an orientation of $G'$ as follows:

- If $v_i$ has out-degree 0, 1 or 3 in $G$, we orient $c_i v_i$ towards $v_i$.

- If $v_i$ has out-degree 2 in $G$, we orient $c_i v_i$ towards $c_i$.

- Let $\mathcal{C}$ be the set of connection vertices $c_i$ in $G'$ that now have an incoming arc $(v_i, c_i)$. Let $g^{\mathcal{C}}$ be the number of connection vertices $c_i \in \mathcal{C}$ where $v_i$ is green. For every $0 \leq i \leq n-1$, the edge $c_i c_{(i+1) \bmod n}$ is directed towards $c_{(i+1) \bmod n}$ if $|\mathcal{C} \cap \{c_0, \ldots, c_i\}| \bmod 2 = g^{\mathcal{C}} \bmod 2$, and towards $c_i$ otherwise.

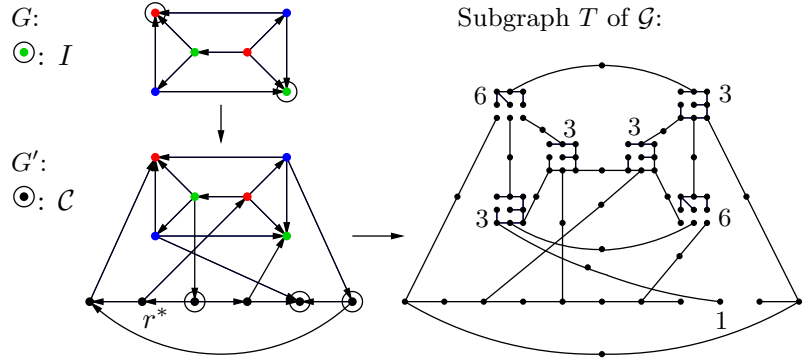

Figure 4: A size 2 independent set yields a spanning connected subgraph with $\lfloor 3.75 \cdot 6 + 1.5 \cdot 2 \rfloor = 25$ weighted leaves.

In Figure 4, $\mathcal{C} = \{c_2, c_4, c_5\}$. $\mathcal{C}$ is represented by encircled vertices of $G'$. Of these vertices, only $c_2$ has a green in-neighbor, so $g^{\mathcal{C}} = 1$. Therefore $c_0 c_1$ is oriented towards $c_0$, etc.

We start with two simple observations on these orientations of $G'$. If a vertex $v_i$ has out-degree 1 in $G$, it retains out-degree 1 in $G'$, and if it has out-degree 2 in $G$ it receives out-degree 3 in $G'$. If it has out-degree 3 in $G$ it retains out-degree 3 in $G'$. This yields:

**Claim 4** *Vertices $v_i$ have out-degree 0, 1 or 3 in $G'$.*

For red vertices $v_i$, either $d_G^+(v_i) = 0$ (if $v_i \in I$), or $d_G^+(v_i) = 3$ (if $v_i \notin I$), so in either case $(c_i, v_i) \in A(G')$. Summarizing:

**Claim 5** *If $v_i$ is red, then $c_i \notin \mathcal{C}$.*

Let $n_d$ denote the number of vertices $v_k$ with $d_G^+(v_k) = d$.

**Claim 6** *$G'$ contains at least $\lfloor n_2/2 \rfloor$ vertices $c_i$ with $d^+(c_i) = 0$.*

*Proof:* Observe that vertices $c_i \in \mathcal{C}$ with $i \geq 1$ have in-degree 1 or in-degree 3 in $G'$, because of the parity based orientation of edges between connection vertices. Recall that there is at least one red vertex, so $v_0$ is red and $c_0 \notin \mathcal{C}$ (Claim 5). Therefore *all* vertices in $\mathcal{C}$ have in-degree 1 or 3, in alternating order of increasing index. Since $|\mathcal{C}| = n_2$, it follows that there are at least $\lfloor n_2/2 \rfloor$ connection vertices with in-degree 3 (and out-degree 0). □

Let $r^* = c_0$ if $g^{\mathcal{C}}$ is even, and $r^* = c_{r-1}$ if it is odd. In Figure 4, $g^{\mathcal{C}} = 1$ so $r^* = c_{r-1} = c_1$.  $r^*$

**Claim 7** *In the chosen orientation of $G'$, every vertex is reachable from $r^*$.*

*Proof:* Out-degrees will refer to $G$ in this proof. First we will show that every vertex $v_i$ of $G'$ is reachable from some connection vertex. If $d_G^+(v_i) \neq 2$, then $v_i$ has a connection vertex as in-neighbor, so the statement is clear. If $d_G^+(v_i) = 2$, then $v_i$ has an in-neighbor $v_x$ in $G'$, with $v_x \notin I$, that must be red or green. If $v_x$ has a connection vertex as in-neighbor, we have proved the statement. Otherwise, $v_x$ has an in-neighbor $v_y$ again, which then must be red. So $v_y$ must have a connection vertex as in-neighbor. In any case, we have found a dipath from some connection vertex to $v_i$.

A connection vertex $c_i$ will be called red, green or blue when its unique (in- or out-) neighbor $v_i$ is red, green or blue respectively. We will now prove that all connection vertices $c_i$ are reachable from $r^*$ in $G'$.

CASE 1: $c_i$ is red.
Since there are no red vertices $c_i \in \mathcal{C}$ (Claim 5), $c_0, c_1, \ldots, c_{r-1}$ is a dipath in $G'$ if $g^{\mathcal{C}}$ is even, and $c_{r-1}, c_{r-2}, \ldots, c_0$ is a dipath if $g^{\mathcal{C}}$ is odd. So we have chosen $r^*$ such that all red connection vertices are reachable from $r^*$.

CASE 2: $c_i \in \mathcal{C}$ is green.
Let $v_i$ be the (green) in-neighbor of $c_i$. The argument we have used above shows that $v_i$ is reachable from some red connection vertex, which in turn is reachable from $r^*$ as shown in case 1.

CASE 3: $c_i \notin \mathcal{C}$ is green.
$c_i$ has a connection vertex as in-neighbor (either $c_{i-1}$ or $c_{i+1}$). If $c_{i-1}$ is its in-neighbor, then $G'$ either contains a dipath $c_{r-1}, \ldots, c_i$, or a dipath $c_j, c_{j+1}, \ldots, c_i$ with $j < i$ and $c_j \in \mathcal{C}$. Both of these dipaths start at a reachable vertex (by case 1 and 2) so $c_i$ is reachable from $r^*$. If $c_{i+1}$ is the in-neighbor of $c_i$, then the number of $\mathcal{C}$ vertices in $\{c_0, \ldots, c_i\}$ has different parity than the number of green vertices in $\mathcal{C}$. Since all $\mathcal{C}$ vertices in $\{c_0, \ldots, c_i\}$ are green (Claim 5), this implies that there is at least one more green vertex in $\mathcal{C}$. So there exists a dipath $c_j, c_{j-1}, \ldots, c_i$ with $j > i$, $c_j$ green, and $c_j \in \mathcal{C}$. $c_j$ is reachable from $r^*$ by case 2, so $c_i$ is reachable as well.

CASE 4: $c_i \in \mathcal{C}$ is blue.
By the same argument as earlier, the blue in-neighbor $v_i$ of $c_i$ is reachable from a red or green connection vertex, which is reachable from $r^*$ by case 1, 2 or 3.

CASE 5: $c_i \notin \mathcal{C}$ is blue.
Similar to the reasoning in case 3, we may trace a path back from $c_i$ consisting of connection vertices, until we find a dipath starting at a vertex $c_j$, where $c_j$ is either red or part of $\mathcal{C}$. (This path may also be $c_0, c_{n-1}, c_{n-2}, \ldots, c_i$, so $j = 0$.) Case 1, 2 and 4 show that $c_j$ and thus $c_i$ is reachable from $r^*$.

Now we have considered all cases for connection vertices. It follows that all vertices of $G'$ are reachable from $r^*$. □

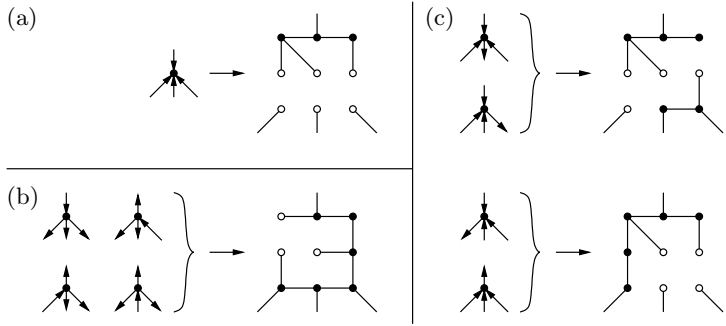

Figure 5: Using out-degrees to construct a spanning tree.

Whenever we refer to the out-degree or in-degree of vertices below, this refers to $G'$, not to $G$, unless explicitly noted otherwise. We use the orientation of $G'$ to construct a spanning tree $T'$ of $\mathcal{G}$ as follows. First we construct a spanning connected subgraph $T$:

1. For every vertex gadget in $\mathcal{G}$, Figure 5 shows which subset of the edges should be chosen in $T$, depending on the out-degree and out-neighbor set of the corresponding vertex $v_i$ in $G'$. (Note that only out-degrees 0, 1 and 3 have to be considered by Claim 4.)

2. Every edge of $\mathcal{G}$ that is not part of a vertex gadget is added to $T$.

3. For every vertex $c_i$ that has in-degree 3 in $G'$, delete the two incident $T$-edges that do not correspond to the arc $(v_i, c_i)$ of $G'$, making $c_i$ a leaf of $T$.

4. Delete edges of $T$ until no cycles remain, to obtain graph $T'$.  $T'$

$T$ denotes the graph as it is after Step 3 above. The following claim already shows for many  $T$
vertices of $\mathcal{G}$ that they are reachable from $r^*$ in $T$.

**Claim 8** *If $G'$ contains a dipath $P' = r^*, \ldots, x, y$ with $d^+(y) \geq 1$, then $T$ contains a path from $r^*$ to $t_{xy}(y)$.*

*Proof:* First, for every arc $(u, v)$ of $P'$ we add the corresponding length 2 path in $\mathcal{G}$ to $P$. To be precise, this is the path $t_{uv}(u), x, t_{uv}(v)$, where $x$ is the vertex resulting from the subdivision of $uv$ during the construction of $\mathcal{G}$. Observe that both of these path edges are also part of $T$: in Step 3 of the construction of $T$ some edges that are not part of vertex gadgets are removed from $T$, but only those that are incident with a vertex $c_i$ with in-degree 3, and thus out-degree 0. Clearly such vertices cannot be internal vertices of $P'$, and by our assumption, the end vertex $y$ of $P'$ also has out-degree at least 1. At this point $P$ may not be a path yet; it can consist of a sequence of paths where one path ends at a terminal $t_1$ of a vertex gadget $H_i$, and the next path starts at another terminal $t_2$ of $H_i$. Joining such paths together is easy when $d^+(v_i) = 3$: Figure 5(b) shows the edges of $T$ that are part of $H_i$; observe that for every terminal pair $t_1$ and $t_2$ a path from $t_1$ to $t_2$ exists in $T$ through $H_i$. So it suffices to prove that $P'$ contains no internal vertices $v_j$ with $d^+(v_j) \neq 3$. Clearly all internal vertices have out-degree at least 1. No vertices $v_j$ of $G'$ have out-degree 2 (Claim 4), so we only have to consider the case that $d^+(v_j) = 1$. Now we will use that we started with a *maximal* independent set $I$: because $I$ is maximal, every vertex that is not in $I$ has at least one neighbor in $I$. So by choice of the orientation of $G$, if $v_j$ has out-degree 1, its out-neighbor $v_k$ is in $I$, and $d^+(v_k) = 0$. The dipath $P'$ cannot contain $v_k$ as internal vertex, and by choice of $P'$, also not as end vertex $y$. Hence $P'$ contains no vertices $v_j$ with out-degree 1. This concludes the proof. □

Using the previous two claims, $T$ can be shown to be connected:

**Claim 9** *All vertices $u \in V(\mathcal{G})$ are reachable from $r^*$ within $T$.*

*Proof:* We consider four cases for $u$.

CASE 1: $u$ is part of a vertex gadget $H_i$, with $d^+(v_i) \geq 1$.
Figure 5(b) and (c) show that in every case, there is an arc $(w, v_i) \in A(G')$ such that $T$ contains a path from $t = t_{wv_i}(v_i)$ to $u$. So we only need to show that $t$ can be reached from $r^*$ within $T$. By Claim 7, $G'$ contains an $(r^*, w)$-dipath, which then yields a dipath $P' = r^*, \ldots, w, v_i$. From Claim 8 it now follows that $T$ contains a $(r^*, u)$-path.

CASE 2: $u$ is part of a vertex gadget $H_i$ with $d^+(v_i) = 0$.
We again consider an arc $(w, v_i) \in A(G')$ such that $T$ contains a path from $t = t_{wv_i}(v_i)$ to $u$ (such an arc exists, see Figure 5(a)). Let $t' = t_{wv_i}(w)$. If $t'$ is part of a vertex gadget, in case 1 we showed that $t'$ can be reached from $r^*$ in $T$, which shows $u$ can be reached. Otherwise, $t' = c_j$ with $d^+(c_j) \geq 1$. Claim 7 shows that $G'$ contains an $(r^*, c_j)$-dipath, which yields an $(r^*, t')$-path in $T$ (Claim 8) and thus an $(r^*, u)$-path.

CASE 3: $u = c_i$.
If $d^+(c_i) \geq 1$, Claim 7 shows that $G'$ contains an $(r^*, c_i)$-dipath, which yields an $(r^*, c_i)$-path in $T$ by Claim 8. If $d^+(c_i) = 0$, then the construction of $T$ shows that both edges of $\mathcal{G}$ corresponding to the arc $(v_i, c_i)$ of $G'$ are part of $T$. By Case 1, every vertex of the vertex gadget $H_i$ corresponding to $v_i$ is reachable from $r^*$ in $T$, so $c_i$ is reachable.

CASE 4: $d(u) = 2$ and $u$ is not part of a vertex gadget.
Here $u$ is the vertex resulting from the subdivision of an edge $xy$. Let $(x, y)$ be the orientation of this edge in $G'$. If $x = c_k$ for some $k$, then Case 3 shows that an $(r^*, c_k)$-path exists in $T$. This can be extended to the desired path; $c_k u \in E(T)$ since $d^+(c_k) \geq 1$. Otherwise, $x \in V(H_i)$, where $d^+(v_i) \geq 1$. Then case 1 or 2 shows that an $(r^*, x)$-path exists in $T$, which can be extended again. □

Since Claim 9 shows that $T$ is connected, clearly $T'$ is connected as well. Since in addition $T'$ contains no cycles, $T'$ is a spanning tree of $\mathcal{G}$. It remains to prove that it has the desired number of leaves. Figure 5 shows that a vertex $v_i$ contributes six leaves to $T$ if $d^+_{G'}(v_i) = 0$, four leaves if $d^+_{G'}(v_i) = 1$ and three leaves if $d^+_{G'}(v_i) = 3$. In addition, every vertex $c_i$ with in-degree 3 in $G'$ is a leaf of $T$ by Step 3 of the construction of $T$. Claim 6 shows that there are at least $\lfloor n_2/2 \rfloor$ such vertices. Recall that $n_d$ denotes the number of vertices that have out-degree $d$ in $G$. In addition let $n'_d$ denote the number of vertices that have out-degree $d$ in $G'$. Observe that $n_0 + n_1 + n_2 + n_3 = n$, and let $m = 1.5n = 3n_0 + 2n_1 + n_2$ be the number of edges of $G$. Together this yields

$$\ell(T) \geq 6n'_0 + 4n'_1 + 3n'_3 + \lfloor n_2/2 \rfloor = 6n_0 + 4n_1 + 3n_2 + 3n_3 + \lfloor n_2/2 \rfloor =$$

$$\lfloor 3n + 3n_0 + n_1 + 0.5n_2 \rfloor = \lfloor 3n + 1.5n_0 + 0.5m \rfloor \geq \lfloor 3.75n + 1.5x \rfloor.$$

For the last step we used that every vertex $u \in I$ has out-degree 0 in $G$ and that $|I| \geq x$. This concludes the proof of Lemma 4.

# 6 Conclusion of the Proof

**Theorem 5** *Cubic MaxLeaf is APX-hard.*

*Proof:* We show that for every $\epsilon > 0$, a $(1 - \epsilon)$-approximation algorithm for cubic MaxLeaf yields a $(1 - 141\epsilon)$-approximation algorithm for Cubic MIS. Let $G$ be a Cubic MIS instance on $n$ vertices, which has a maximum independent set of size $x$. Observe that since $G$ is cubic, $x \geq n/4$. From $G$, we construct a Weighted MaxLeaf instance $\mathcal{G}$ as shown in Section 3. $\mathcal{G}$ has a tree with at least $\lfloor 3.75n + 1.5x \rfloor$ weighted leaves (Lemma 4), and it can be checked that it has $y = 4.5n$ vertices of degree 2. Let $r = 3.75n + 1.5x - \lfloor 3.75n + 1.5x \rfloor$. Note that since $n$ is even, the rounded value is half-integral so $r \leq 0.5$. From $\mathcal{G}$, we construct a Cubic MaxLeaf instance $H$ by replacing degree 2 vertices as shown in Section 3. Then $H$ has a tree with at least $3.75n + 1.5x - r + 3y = 3.75n + 1.5x - r + 13.5n$ leaves (Lemma 1).

Now suppose we have a $(1 - \epsilon)$-approximation algorithm for cubic MaxLeaf. In $H$, this algorithm will find a tree $T$ with at least $(1-\epsilon)(3.75n+1.5x-r+13.5n)$ leaves. By Lemma 1 again, this yields tree $T'$ of $\mathcal{G}$ with at least $(1-\epsilon)(3.75n+1.5x-r+13.5n) - 13.5n$ weighted leaves. So, using $x \geq n/4$, we obtain:

$$\ell(T') \geq 3.75n + 1.5x - r - \epsilon(3.75n + 1.5x - r + 13.5n) =$$

$$3.75n + 1.5x - r - \epsilon(17.25n + 1.5x - r) \geq$$

$$3.75n + 1.5x - r - \epsilon(69x + 1.5x) = 3.75n + 1.5x - r - \gamma x,$$

where $\gamma = 70.5\epsilon$. Now we consider two cases:

If $\gamma x < 0.5$, then $\ell(T') > 3.75n + 1.5x - 0.5 - 0.5 = 3.75n + 1.5(x - \frac{2}{3})$. (Here we used $r \leq 0.5$.) By Lemma 3, we can construct an independent set $I$ for $G$ with $|I| > x - \frac{2}{3} - \frac{1}{3}$ (note that the inequality is again strict). $x$ is integer, so $|I| \geq x$. Hence in this case we find an optimal independent set.

On the other hand, if $\gamma x \geq 0.5$, then also $\gamma x \geq r$, so $\ell(T') > 3.75n + 1.5x - \gamma x - \gamma x = 3.75n + 1.5(x - \frac{4}{3}\gamma x)$. So by Lemma 3 again, we find $I$ with $|I| \geq x - \frac{4}{3}\gamma x - \frac{1}{3} \geq x - 2\gamma x$. In this case we have an $(1 - 2\gamma) = (1 - 141\epsilon)$ approximation. Since Cubic MIS is APX-hard [1], the APX-hardness of Cubic MaxLeaf follows. □

We remark that this reduction is an L-reduction as introduced in [21]. Similarly, using the fact that cubic graphs on $n$ vertices have a spanning with at least $n/4 + 2$ leaves [15], we find that a $(1 + \epsilon)$-approximation algorithm for MinCDS yields a $(1 - 3\epsilon)$-approximation algorithm for Cubic MaxLeaf on the same graph, so:

**Corollary 6** *Cubic MinCDS is APX-hard.*

*Proof:* We consider the trivial reduction from cubic MaxLeaf. Let $G$ be a cubic graph on $n$ vertices for which we wish to find a spanning tree with maximum number of leaves. Let $l$ be the maximum number of leaves possible for $G$. Since $G$ is cubic, $l \geq n/4 + 2$ [15].

$G$ then has a connected dominating set of size at most $n - l$. A $(1 + \epsilon)$-approximation algorithm for MinCDS returns a solution $S$ with

$$|S| \leq (1 + \epsilon)(n - l) = n - l - \epsilon l + \epsilon n < n - l - \epsilon l + 4\epsilon l = n - l + 3\epsilon l.$$

So $S$ can be used to find in polynomial time a spanning tree with at least $l - 3\epsilon l$ leaves, which together yields a $(1 - 3\epsilon)$-approximation algorithm for cubic MaxLeaf. The APX-hardness of Cubic MinCDS follows. □